\documentclass[twocolumn,nofootinbib,superscriptaddress]{revtex4-1}
\pdfoutput=1
\usepackage{epsfig}
\usepackage{calligra,amsmath,amsfonts,mathrsfs,amssymb,bbm,textcomp,bbding}
\usepackage{slashed,cancel,units}
\usepackage[usenames,dvipsnames]{xcolor}
\usepackage{hyperref}
\hypersetup{colorlinks=true,urlcolor=Magenta,anchorcolor=blue,citecolor=blue,filecolor=blue,
            linkcolor=Magenta,menucolor=blue, linktocpage=true,pdfproducer=medialab}
\usepackage[utf8]{inputenc}
\usepackage{bm} 

\newcommand{\blue}[1]{\color{blue} #1 \color{black}}

\definecolor{vierde}{rgb}{0.0, 0.5, 0.0}

\definecolor{OliveGreen}{rgb}{0,0.6,0}

\newcommand{\be}{\begin{equation}}
\newcommand{\ee}{\end{equation}}
\newcommand{\ba} {\begin{equation}\begin{aligned}}
\newcommand{\ea} {\end{aligned}\end{equation}}
\newcommand{\bea}{\begin{eqnarray}}
\newcommand{\eea}{\end{eqnarray}}

\newcommand{\LL}{\mathscr{L}}

\newcommand{\OO}{\mathcal{O}}

\newcommand{\absval}[1]{\left| #1 \right|}

\newcommand{\hc}{\text{h.c.}}
\newcommand{\ov}[1]{\overline{#1}}
\newcommand{\nn}{\nonumber}

\newcommand{\GeV}{\ \text{GeV}}

\def\diag{{\tt diag}}

\begin{document}
\eprint{FTUAM-21-2
\hfill IFT-UAM/CSIC-21-29}

\title{\bf\LARGE \blue{
A New bound on CP Violation in the $\tau$ Lepton Yukawa Coupling and electroweak baryogenesis}}

\author{\bf J.~Alonso-Gonz\'alez}
\email{j.alonso.gonzalez@csic.es}
\affiliation{Instituto de F\'isica Te\'orica UAM/CSIC, Calle Nicol\'as Cabrera 13-15, Cantoblanco E-28049 Madrid, Spain}
\affiliation{Departamento  de  F\'{\i}sica Te\'{o}rica,  Universidad  Aut\'{o}noma  de  Madrid, Cantoblanco  E-28049  Madrid,  Spain}

\author{\bf L.~Merlo}
\email{luca.merlo@uam.es}
\affiliation{Instituto de F\'isica Te\'orica UAM/CSIC, Calle Nicol\'as Cabrera 13-15, Cantoblanco E-28049 Madrid, Spain}
\affiliation{Departamento  de  F\'{\i}sica Te\'{o}rica,  Universidad  Aut\'{o}noma  de  Madrid, Cantoblanco  E-28049  Madrid,  Spain}

\author{\bf S.~Pokorski}
\email{Stefan.Pokorski@fuw.edu.pl}
\affiliation{Institute of Theoretical Physics, Faculty of Physics, University of Warsaw, Pasteura 5, PL 02-093, Warsaw, Poland}
\begin{abstract}
The origin of the matter-antimatter asymmetry in the Universe is a fundamental question of physics. Electroweak baryogenesis is a compelling scenario for explaining it but it requires beyond the Standard Model sources of the CP symmetry violation. The simplest possibility is CP violation in the third generation fermion Higgs couplings, widely investigated theoretically and searched for experimentally. It has been found that the experimental bounds on the CP violation in the quark Yukawa couplings exclude their significant role in the electroweak baryogenesis, but it can be still played by the $\tau$ lepton Yukawa coupling. It is shown in this paper that, within the context of the Standard Model Effective Field Theory and assuming an underlying flavour symmetry of the Wilson coefficients, the electron dipole moment bound on the $\tau$ lepton Yukawa coupling is two orders of magnitude stronger than previously reported. This sheds strong doubts on its role in the electroweak baryogenesis, further stimulates the interest in its experimental verification and makes electroweak baryogenesis even more difficult to explain.
\end{abstract}

\maketitle

\section{Introduction}
\label{sec:intro}

CP violating third generation Yukawa couplings would be an important source of baryon-antibaryon asymmetry in the electroweak baryogenesis~\cite{deVries:2017ncy,Kobakhidze:2015xlz,Guo:2016ixx,Chiang:2016vgf,Fuyuto:2017ewj,deVries:2018tgs,Fuchs:2020uoc,Fuchs:2020pun,Xie:2020wzn}. However, there is certain tension between their magnitude required by the observed baryon asymmetry in the Universe (BAU) and the bounds following mainly from the experimental upper limit on the electron Electric Dipole Moment (EDM)~\cite{Andreev:2018ayy}. The Higgs fermion effective Lagrangian, below the electroweak symmetry breaking (EWSB) scale and in the mass eigenstate basis, can be written as~\cite{Brod:2013cka}
\be
\LL_\text{eff}=-\dfrac{y_\psi}{\sqrt2}\left(\kappa_\psi \ov{\psi}\psi+i\tilde{\kappa}_{\psi}\ov\psi\gamma_5\psi\right)h\,,
\label{Brod_Lag}
\ee
where $\psi$ refers to either quarks or charged leptons with $y_\psi=\sqrt2 m_\psi/v$, $m_\psi$ is the $\psi$ mass and  $v=246\GeV$ is the EW symmetry breaking vacuum expectation value (VEV) of the Higgs field. The $\kappa_\psi$ and $\tilde \kappa_\psi$ are real numbers parametrising the CP conserving  and CP violating parts of these couplings. In the SM, $\kappa_\psi=1$ and $\tilde \kappa_\psi=0$. With some beyond the Standard Model (SM) CP violation effects encoded in non-vanishing parameters $\tilde \kappa_\psi$, one obtains contributions to the electron EDM from the Barr-Zee diagrams with the fermions $\psi$ running in the loop.
Most recent studies such as Ref.~\cite{Fuchs:2020uoc}, assuming only one third generation fermion running in the loop at a time~\footnote {Those bounds are slightly relaxed when contributions from several fermions in the loop are included.}, give:
\be
|\tilde\kappa_t|\lesssim 0.0011\,,\quad
|\tilde\kappa_b|\lesssim 0.25\,,\quad
|\tilde\kappa_\tau|\lesssim 0.3\,.
\label{Boundkappatildetau1}
\ee
These bounds are usually interpreted in the framework of the SM Effective Field Theory (SMEFT)~\cite{Buchmuller:1985jz,Grzadkowski:2010es}, with  $d=6$ operators with complex Wilson coefficients contributing to the third generation Yukawa couplings added to the SM Lagrangian~\cite{deVries:2017ncy,deVries:2018tgs,Fuchs:2020uoc}. 
The  bounds on the CP violation in the top and bottom couplings are by far too strong for them to play a significant role in the EW baryogenesis (because of the strong sphaleron effects) but there is still enough room for CP violation in the $\tau$ Yukawa coupling (see also Refs.~\cite{Joyce:1994bi,Chung:2009cb,Chung:2008aya} in the frameworks of two Higgs doublets model and the minimal supersymmetric SM). Indeed, provided the first order phase transition is strong enough, the smallest value for $\tilde\kappa_\tau$ to entirely explain BAU, and  being compatible with LHC Higgs signal strength at $2\,\sigma$, reads~\cite{Fuchs:2020uoc} (see projected sensitivities at future lepton colliders for CPV $h\tau\tau$ coupling in~\cite{Ge:2020mcl})
\be
|\tilde\kappa_\tau| \gtrsim 0.08\,.
\label{LowerTildeKTau}
\ee

In this paper we emphasise the fact that in the EFT framework to beyond the SM physics, under such assumptions like Minimal Flavour Violation (MFV)~\cite{Chivukula:1987py,DAmbrosio:2002vsn,Cirigliano:2005ck,Davidson:2006bd,Alonso:2011jd} or flavour symmetries underlying the fermion mass hierarchies and mixings, the Wilson coefficients of the $d=6$ operators have in general certain flavour structure.
In that case the imaginary parts of the $h\tau\tau$ and $hee$ couplings are linked to each other.  We point out that
the Barr-Zee contribution to the electron EDM gives very strong bound on $\tilde\kappa_e$
\be
|\tilde\kappa_e|\lesssim 0.0017\,,
\label{Boundkappatildee}
\ee
which, under the above mentioned flavour structures, leads to the bound on $\tilde\kappa_\tau$ two orders of magnitude stronger than the one from the $\tau$-loop  contribution in the Barr-Zee diagram:
\be
|\tilde\kappa_\tau|\lesssim \alpha\,0.0017\,,
\label{Boundkappatildetau2}
\ee
where $\alpha$ is $\OO(1)$. As we will discuss later, this result holds also in the case in which more than one contribution to the electron EDM is taken into consideration. This bound is incompatible with the one in Eq.~\eqref{LowerTildeKTau}, thus making highly unlikely the possibility to successfully explain the current value of BAU with  new sources of CPV only  in the tau lepton Yukawa coupling. 

In the following, in Sect.~\ref{sec:EDM}, we first recall the general formulae for the electron EDM, while in Sect.~\ref{sec:Examples} we will provide a series of examples. Finally, we conclude in Sect.~\ref{sec:concl}.

\section{The Electron EDM}
\label{sec:EDM}

The effective EDM Lagrangian for the electron can be written as
\be
\LL_\text{EDM}=-d_e\frac{i}{2}\ov{e}\sigma^{\mu\nu}\gamma_5 e F_{\mu\nu}
\ee
where $d_e$ is the EDM coefficient, $e$ the electron field, $\sigma^{\mu\nu}$ the antisymmetric two-dimensional tensor, and $F_{\mu\nu}$ the electromagnetic gauge field tensor. The most recent upper bound on the electron EDM is from the ACME II collaboration~\cite{Andreev:2018ayy} and reads 
\begin{equation}
|d_e|<1.1\times 10^{-29}\text{ e cm} \, , \qquad  \text{at }90\%\text{ C.L.} \, .
\label{BoundeEDM}
\end{equation}

\begin{figure}[h!]
\centering
\includegraphics[width=0.49\textwidth]{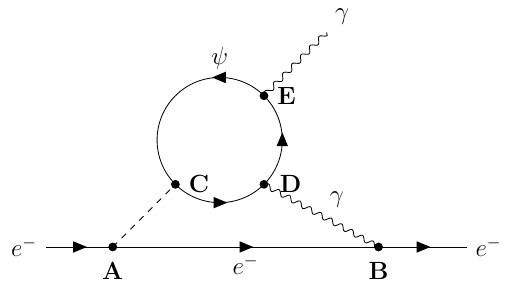} 
\caption{\em Two-loop Barr-Zee diagram for the electron EDM.}
\label{fig:BZ}
\end{figure}

The electron EDM  receives contributions  from the so-called Barr-Zee diagram  shown in Fig.~\ref{fig:BZ}. We assume that the photon-fermion  vertices  $B$, $D$ and $E$ are  as in the SM (in particular, they are left unchanged by the $d=6$ operators of SMEFT) and calculate the bounds on $\tilde\kappa_\tau$ and $\tilde\kappa_e$  introduced in Eq.~\eqref{Brod_Lag}, following from the experimental limit Eq.~\eqref{BoundeEDM}.

The explicit computation of the Barr-Zee diagram  gives~\cite{Brod:2013cka}
\be
\begin{split}
\frac{d_e}{e}=&4 \, N_C \, Q^2_\psi \, \dfrac{\alpha_\text{em}}{(4\pi)^3} \, \sqrt2  \, G_F \, m_e\times \\
&\hspace{0.5cm}\times\left[\kappa_e\tilde\kappa_\psi f_1(x_{\psi/h})+\tilde\kappa_e\kappa_\psi f_2(x_{\psi/h})\right]
\end{split}
\label{GenericEDM}
\ee
where $Q_\psi$ is the $\psi$ electric charge, $N_C=3$ if the fermion $\psi$ running in the loop is a quark and $N_C=1$ if it is a charged lepton, $\alpha_\text{em}$ is the fine structure constant at the scale of the electron mass, $G_F$ the Fermi constant, and $f_1$ and $f_2$ are functions of $x_{\psi/h}=m_\psi^2/m_h^2$, defined by
\be
\begin{aligned}
f_1(x)&=\dfrac{2x}{\sqrt{1-4x}}\left[Li_2\Bigg(1-\dfrac{1-\sqrt{1-4x}}{2x}\right)+\\
&\hspace{1.82cm}-Li_2\left(1-\dfrac{1+\sqrt{1-4x}}{2x}\right)\Bigg]\\
f_2(x)&=(1-2x)f_1(x)+2x(\ln x+2)\,,
\end{aligned}
\ee
where $Li_2$  is the usual dilogarithm
\be
Li_2(x)=-\int_0^x du \dfrac{\ln(1-u)}{u}\,.
\ee

In the leading order in new physics effects, the bound on $\tilde\kappa_\tau$ is obtained when the lepton $\tau$ is running in the loop and for $\tilde\kappa_e =0 $. One gets
\be
\absval{\tilde\kappa_\tau}\lesssim \,0.3\,.
\label{weakbound}
\ee

The bound on $\tilde\kappa_e$ is obtained for the top quark running in the loop and for $\tilde\kappa_t =0$. We get
\be
\absval{\tilde\kappa_e}\lesssim 0.3\left(\frac{3}{4}\right)\frac{f_1(x_{\tau/h})}{f_2(x_{t/h})}=0.0017\,.
\label{strongbound}
\ee
We see that the bound on $\tilde\kappa_e$ is more than two orders of magnitude stronger than on $\tilde\kappa_\tau$. In the  following  we show that under  very general assumptions, $\tilde\kappa_e$ and $\tilde\kappa_\tau $ are linked to each other  and the bound in Eq.~\eqref{strongbound}  gives a bound on $\tilde\kappa_\tau $ about two orders of magnitude stronger than the one in Eq.~\eqref{weakbound}.

We assume that the dominant contribution to the effective lepton Yukawa couplings, Eq.~\eqref{Brod_Lag}, comes from  $d=4$ and $d=6$ operators of the SMEFT Lagrangian.  Furthermore, we assume that the flavour structure of that Lagrangian is either compatible with  the MFV hypothesis or controlled by a flavour symmetry that is responsible for the charged lepton masses. The Yukawa Lagrangian then reads \footnote{Notice that the $d=6$ operator introduced in Eq.~\eqref{Non-RenormLag} may give rise to potentially interesting FCNC effects~\cite{AMP}.\label{fn:FCNC}}
\be
\LL=
-\ov{L'} H Y' e'_R
-\ov{L'} H C' e'_R\dfrac{H^\dag H}{\Lambda^2}+\hc\,,
\label{Non-RenormLag}
\ee
where the prime denotes  fields and quantities in the flavour basis (the flavour indices are omitted). $L'$ are the lepton EW doublets, $e'_R$ stands for the EW singlets and  $H$ is the Higgs EW doublet. The scale $\Lambda$ is the effective scale  of new physics. Finally $Y'$ and $C'$ are  $3\times 3$ complex matrices in the flavour space, controlled by the symmetry assumption. We have also assumed,  and this is our third assumption, that the scale $\Lambda$ is universal for all terms.

Below the scale of the EWSB we have
\begin{align}
\LL=
&-\ov{e'_L} \left(Y'+\dfrac{v^2}{2\Lambda^2}C'\right) e'_R\dfrac{v}{\sqrt2}+
\label{LagH}\\
&-\ov{e'_L} \left(Y'+\dfrac{3v^2}{2\Lambda^2}C'\right) e'_R\dfrac{h}{\sqrt2}+\hc+\ldots\,,\nn
\end{align}
where the first  term gives the fermion masses, while the second term describes the fermion interactions with the physical Higgs field. Dots stand for terms with more than one Higgs field.

The Yukawa matrices are diagonalised by the rotations on the left-  and right-handed fields:
\be
Y'+\dfrac{v^2}{2\Lambda^2}C'=VYU^\dag
\label{Rotations}
\ee
where  $V$ and $U$ are unitary $3\times3$ matrices. The matrix   $V$ enters into the definition of the PMNS matrix. The matrix $Y$ is a diagonal matrix with the  entries $Y=\sqrt2/v \,\left(m_e,m_\mu, m_\tau\right)$. The Lagrangian in the mass eigenstate basis  then reads:
\be
\LL=
-\ov{e_L} Y e_R\dfrac{v}{\sqrt2}
-\ov{e_L} \left(Y +\dfrac{v^2}{\Lambda^2}C\right) e_R\dfrac{h}{\sqrt2}+\hc+\ldots\,,
\label{LagHMassBasis}
\ee
where
\be
C= V^\dag C'U\,.
\label {CC'}
\ee

The matching between the two effective Lagrangians, Eqs.~\eqref{Brod_Lag} and \eqref{LagHMassBasis},  gives
\be
\begin{aligned}
Y K=&Y + \dfrac{v^2}{\Lambda^2}\diag(\texttt{Re} C)\\
Y{\tilde K}=&\dfrac{v^2}{\Lambda^2}\diag(\texttt{Im} C)\,,
\end{aligned}
\label{Kappa_matrices}
\ee
where $ K$ is the diagonal matrix  containing the $\kappa_\psi$ parameters for the charged leptons, \mbox{$ K\equiv\diag(\kappa_e,\,\kappa_\mu,\,\kappa_\tau)$}.
Thus, in particular, we get
\begin{align}
\tilde\kappa_\tau=&\dfrac{v^2}{\Lambda^2}\dfrac{\texttt{Im}~ C_{33}}{y_\tau}
\label{KappaCe33C}\\
\tilde\kappa_e=&\dfrac{v^2}{\Lambda^2}\dfrac{\texttt{Im}~ C_{11}}{y_e}\,.
\label{KappaCe11C}
\end{align}
Therefore, the electron EDM bound on $\tilde\kappa_e $ leads to  the following bound on $\tilde\kappa_\tau$:
\be
\absval{\tilde\kappa_\tau}\lesssim 0.0017 \, \dfrac{m_e}{m_\tau} \absval{\dfrac{\texttt{Im} ~C_{33}}{\texttt{Im}~C_{11}}}\,,
\label{FinalFormulaSupGen}
\ee
where  the matrix $C$ is given  in terms of the original Lagrangian parameters by the matrix equation in Eq.~\eqref{CC'}:
\be
C_{ii}=V^*_{ki}C'_{kl}U_{li} \, .
\label{CC' matrix}
\ee

In the following section we discuss the significance of the bound in Eq.~\eqref{FinalFormulaSupGen}  for several concrete scenarios of the flavour  structure of the SMEFT Yukawa Lagrangian.

\section{Concrete Scenarios}
\label{sec:Examples}

We first consider the case when the flavour structure of the Lagrangian in Eq.~\eqref{Non-RenormLag} is compatible with the MFV hypothesis~\cite{Chivukula:1987py,DAmbrosio:2002vsn,Cirigliano:2005ck,Davidson:2006bd,Alonso:2011jd}\footnote{The flavour symmetry usually associated to the MFV framework is a product of $U(3)$ terms. In this context, new physics with flavour violating effects are expected to be heavier than a few TeV~\cite{Chivukula:1987py,DAmbrosio:2002vsn,Cirigliano:2005ck,Davidson:2006bd,Alonso:2011jd,Cirigliano:2006su,Grinstein:2006cg,Paradisi:2009ey,Grinstein:2010ve,Feldmann:2010yp,Guadagnoli:2011id,Buras:2011zb,Buras:2011wi,Feldmann:2009dc,Alonso:2011yg,Alonso:2012fy,Alonso:2013mca,Alonso:2013nca,Dinh:2017smk,Arias-Aragon:2017eww,Merlo:2018rin,Arias-Aragon:2020qip}.}.  Then, in the electroweak basis
\be
C'=c'Y'
\label{MFV}
\ee
where $c'$ is a flavour blind complex number.  It follows from Eq.~\eqref{FinalFormulaSupGen} that:
\be
\absval{\tilde\kappa_\tau}\lesssim0.0017\,.
\label{MFVbound}
\ee
The same result holds for models based on slightly different flavour symmetries than the MFV, but with very similar construction, such as the data driven flavour model~\cite{Arias-Aragon:2020bzy} and models based on $U(2)^n$ flavour symmetry~\cite{Barbieri:2011ci,Blankenburg:2012nx}. As a confirmation of our results, one can see a similar bound was obtained in Ref.~\cite{Egana-Ugrinovic:2018fpy} with MFV in a type III two Higgs Doublet Model scenario for large $\tan(\beta)$.

This bound is two orders of magnitude stronger than the bound on $\tilde\kappa_\tau$ obtained from the contribution to the electron EDM from the Barr-Zee diagram with the $\tau$-loop and it follows from the assumption about the MFV-like flavour structure of the Lagrangian in Eq.~\eqref{Non-RenormLag}. In particular, the bound Eq.~\eqref{MFVbound} applies also when the Yukawa matrix is already diagonal in the electroweak basis.

Our second example are Froggatt-Nielsen models~\cite{Froggatt:1978nt} with horizontal $U(1)$ symmetry.
The $U(1)$ invariant Yukawa terms are given by non-renormalisable operators constructed with fermion bilinears, the SM Higgs doublet and powers of the additional scalar $\phi$, singlet under the SM gauge symmetries, suppressed by the cut-off scale $\Lambda_F$. The latter is different from the scale $\Lambda$ that suppresses the $d=6$ operators in Eq.~\eqref{Non-RenormLag}: the two scales indeed correspond to two different sectors of the theory and have, in general, nothing in common. 

Taking for concreteness, but without any loss of generality, the $U(1)$-charges of the fermion fields $\ov{L_{i}}$ ($e_{Ri}$) as positive integers $n_{Li}$ ($n_{Ri}$) and the $U(1)$-charge for the scalar $\phi$ as $n_\phi=-1$, for the Lagrangian  Eq.~\eqref{Non-RenormLag} we get:
\begin{align}
\LL_\text{FN}=&
-y'_{ij}\ov{L'_{i}} H  e'_{Rj}\left(\dfrac{\phi}{\Lambda_F}\right)^{(n_{Li}+n_{R_j})}+
\label{Non-RenormLagFN}\\
&-c'_{ij}\ov{L'_{i}} H e'_{Rj}\dfrac{H^\dag H}{\Lambda^2}\left(\dfrac{\phi}{\Lambda_F}\right)^{(n_{Li}+n_{R_j})}+\hc\,,\nn
\end{align}
where $y'_{ij}$ and $c'_{ij}$ are free complex parameter with moduli of order $\OO(1)$. The different operators are suppressed by the powers of  the ratio $\epsilon = \langle\phi\rangle/\Lambda_F$, that depend on the fermion charges. Typically $n_{L1}>n_{L2}>n_{L3}$ and $n_{R1}>n_{R2}>n_{R3}$: this guarantees the correct charged lepton mass hierarchy and small mixings. \footnote{For a discussion of various constraints on the charge assignment in the Froggatt-Nielsen models see, for instance,  Refs.~\cite{Dudas:1995yu,Chankowski:2005qp,Altarelli:2002sg,Altarelli:2012ia,Bergstrom:2014owa} and references therein.} The latter is not a very strong requirement, but follows from the prejudice that the large mixings of the PMNS matrix arise mainly from the neutrino sector. This condition on the charges guarantees a simple parametrisation of the mixing angles. According to the definition in Eq.~\eqref{Rotations}:
\begin{align}
Y=&\diag(y_e\,\epsilon^{n_{L1}+n_{R1}},\,y_\mu\,\epsilon^{n_{L2}+n_{R2}},\,y_\tau\,\epsilon^{n_{L3}+n_{R3}})\nn\\
V_{ij}=&\delta_{ij} + (1-\delta_{ij})\frac{(j-i)}{|j-i|}\frac{y_{ij}}{y_{jj}}\epsilon^{\absval{n_{Li}-n_{Lj}}}\\
U_{ij}=&\delta_{ij} + (1-\delta_{ij})\frac{(j-i)}{|j-i|}\frac{y_{ji}}{y_{jj}}\epsilon^{	\absval{n_{Ri}-n_{Rj}}}\,,\nn
\end{align}
where $y_{ij}$ and $y_i$ are complex and real parameters of $\OO(1)$, respectively. It follows from Eq.~\eqref{CC' matrix}  that in $C_{33}$ there is only one dominant term,
\be
C_{33}\simeq \OO(c'_{33})\epsilon^{n_{L3}+n_{R3}}\simeq \OO\left(\dfrac{\sqrt2m_\tau}{v}\right)e^{i\theta_1}\,,
\label{FNC33}
\ee
whereas all the terms entering in the definition of $C_{11}$ are of the same order of magnitude,
\be
C_{11}\simeq \OO(c'_{11})\epsilon^{n_{L1}+n_{R1}}\simeq \OO\left(\dfrac{\sqrt2m_e}{v}\right)e^{i\theta_2}\,.
\label{FNC11}
\ee
In the last equations, we have introduced effective phases, $\theta_1$ and $\theta_2$, for the sum of terms in Eq.~\eqref{CC' matrix}.

\begin{figure}[h!]
\centering
\includegraphics[width=0.49\textwidth]{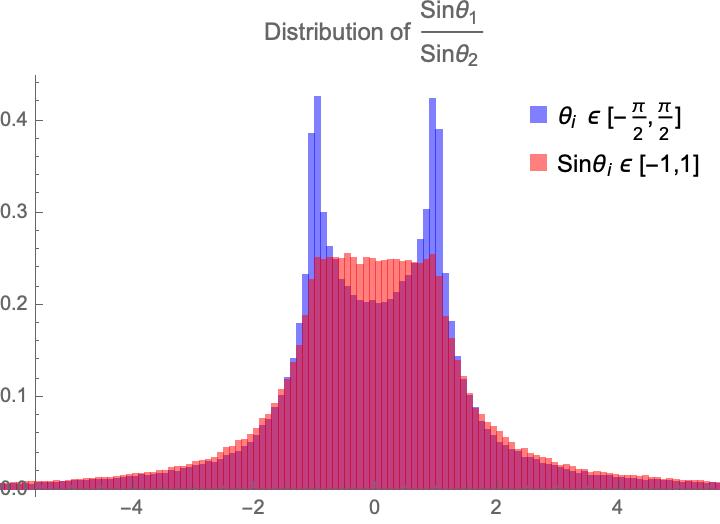} 
\caption{\em Probability distribution of the ratio $\sin\theta_1/\sin\theta_2$. In blue the $\theta_i$ angles are taken randomly linearly distributed in the interval $[-\pi/2,\pi/2]$; in red the $\sin\theta_i$ are taken randomly linearly distributed in the interval $[-1,1]$.}
\label{fig:RatioSines}
\end{figure}

As the final result, using Eqs.~\eqref{FinalFormulaSupGen}, \eqref{FNC33} and \eqref{FNC11}, we get:
\be
\absval{\tilde\kappa_\tau}\lesssim 0.0017 \, \absval{\dfrac{\sin\theta_1}{\sin\theta_2}}\OO(1)\,.
\label{FinalFormulaGenFN}
\ee
The bound in Eq.~\eqref{FinalFormulaGenFN} depends on the phases $\theta_i$, which are not fixed by the U(1) symmetry.   Despite it, the ratio of the two sines has a well determined statistical distribution as it can be seen in Fig.~\ref{fig:RatioSines}: the probability distribution depicted in blue refers to the case in which the angles $\theta_1$ and $\theta_2$ are taken to be randomly linearly distributed in the interval $[-\pi/2,\pi/2]$ and the ratio of the sines are picked at $\pm1$; in red  we show the case in which the sines of the angles are randomly linearly distributed in the interval $[-1,1]$ and in this case all the region between $-1$ and $1$ is uniformly filled in. This naive estimation of the ratio of the sines leads to the conclusion that, while an enhancement may occasionally occur, it is statistically very unlikely.

A third concrete case we consider is the one of discrete flavour models~\cite{Altarelli:2010gt,Altarelli:2012ss,Bazzocchi:2012st,King:2013eh,King:2015aea}, proposed several years ago to obtain the so-called Tri-Bimaximal mixing texture~\cite{Harrison:2002er,Xing:2002sw} as a good description of the PMNS mixing matrix. Its main features are a maximal atmospheric angle, a vanishing reactor angle and a solar angle close to the corresponding experimental central value. In the vast majority of models, the charged lepton mass matrix at leading order is diagonal, while the neutrino mass matrix gives rise to a PMNS with the Tri-Bimaximal texture. Next-to-leading order (NLO) corrections are present and necessary in order to explain a non-vanishing reactor angle and deviation from the maximal value for the atmospheric angle in agreement with the data. As discussed in the reviews~\cite{Altarelli:2012bn,Altarelli:2012ss,Bazzocchi:2012st}, the charged lepton mass matrix at NLO in the LR basis looks like 
\be
M'=m_j\delta_{ij}+ y'_{ij} m_j \epsilon
\ee
where $m_j$ are the charged lepton masses at LO, $\delta_{ij}$ is the Dirac delta matrix, $y'_{ij}$ are free complex parameter with moduli of order $\OO(1)$, and $\epsilon$ is a parameter smaller than 1 that indicates the relative suppression of the higher-order contributions. The $C'$ matrix has the same flavour structure as $M'$. 

The reactor angle at NLO is predicted to be proportional to $\epsilon$ and therefore it is possible to fix the latter equal to $\epsilon\simeq\OO(\sin\theta_{13}^\text{exp})\simeq\OO(0.15)$. It follows that the dominant term in $C_{33}$ is still $C'_{33}\propto m_\tau$, while the dominant one in $C_{11}$ is still $C'_{11}\propto m_e$. We can conclude that the bound in this class of flavour models on $\tilde\kappa_\tau$ turns out to be again the one shown in Eq.~\eqref{FinalFormulaGenFN}, where $\OO(1)$ stands for the unknown coefficients of the NLO corrections. 

We can also comment on models where the flavour symmetry is a continuous non-Abelian one. Without entering into details of a specific model, we can safely state that the final result for the bound on $\tilde\kappa_\tau$ interpolate between the expression for the MFV case in Eq.~\eqref{MFVbound} and the one for the FN or discrete symmetry case in Eq.~\eqref{FinalFormulaGenFN}.

\section{Final Remarks}
\label{sec:concl}

We have shown that the strong experimental bound from the electron EDM on the imaginary part of the $hee$ coupling can be translated into a bound on  the imaginary part of $h\tau\tau$ coupling whenever the MFV ansatz or similar assumptions (such as diagonal corrections to the Yukawa couplings) or flavour symmetries are acting in the flavour sector. This bound is about two order of magnitude stronger than the one obtained from   the $\tau$-loop contribution to the Barr-Zee diagram. While we obtained this conclusion assuming single new physics contribution to the electron EDM at a time, it can be safely generalised. Indeed, the generic expression for $d_e/e$ from Eq.~\eqref{GenericEDM} reads
\be
\dfrac{d_e}{e}=4\, \dfrac{\alpha_\text{em}}{(4\pi)^3} \, \sqrt2  \, G_F \, m_e\times \tilde\kappa^\text{eff}
\ee
where
\be
\tilde\kappa^{eff}=\left[2.68\tilde\kappa_e +3.83\tilde\kappa_t +0.018\tilde\kappa_b +0.015\tilde\kappa_\tau \right]
\ee
Given  the experimental bound  $\absval{\tilde\kappa^{eff}}<0.0045$, we see that the saturation of the bound on $\tilde\kappa_\tau$ from Eq.~\eqref{Boundkappatildetau1} would require 1:1000 cancellation between $\tilde\kappa_e$ and $\tilde\kappa_t$, if Eq.~\eqref{KappaCe33C}  and Eq.~\eqref{KappaCe11C} are to be  preserved.

Our results have an impact on the possibility of explaining the correct amount of BAU: indeed we  have shown that under the assumption of typical flavour scenarios for the Wilson coefficients of higher dimension contribution to the  lepton Yukawa couplings, the electron EDM bound excludes the values of the parameter $\tilde\kappa_\tau$ necessary for explaining the current asymmetry.

\section*{Acknowledgements}

J.A.G. and L.M. acknowledge partial financial support by the Spanish MINECO through the Centro de excelencia Severo Ochoa Program under grant SEV-2016-0597, by the Spanish ``Agencia Estatal de Investigac\'ion''(AEI) and the EU ``Fondo Europeo de Desarrollo Regional'' (FEDER) through the project PID2019-108892RB-I00/AEI/10.13039/501100011033, by the European Union's Horizon 2020 research and innovation programme under the Marie Sk\l odowska-Curie grant agreement No 860881-HIDDeN. L.M. acknowledges partial financial support by the Spanish MINECO through the ``Ram\'on y Cajal'' programme (RYC-2015-17173). The research of S.P. has received funding from the Norwegian Financial Mechanism for years 2014-2021, grant nr 2019/34/H/ST2/00707. L.M. thanks Enrique Fernandez Martinez for discussions. S.P. thanks Anna Lipniacka (University of Bergen) for discussions on the experimental aspects of measuring the imaginary part of the $h\tau\tau$ coupling.

\footnotesize

\bibliography{biblio.bib}
\bibliographystyle{BiblioStyle.bst}

\end{document}